\documentclass[12pt]{article}
\usepackage{cite,epsfig,amssymb,amsmath,graphicx,color,fleqn}

\newcommand{\RNum}[1]{\uppercase\expandafter{\romannumeral #1\relax}}

\newcommand{\be}{\begin{equation}}
\newcommand{\ee}{\end{equation}}
\newcommand{\bear}{\begin{eqnarray}}
\newcommand{\eear}{\end{eqnarray}}
\newcommand{\ba}{\begin{array}}
\newcommand{\ea}{\end{array}}
\newcommand{\nn}{\nonumber}

\usepackage[normalem]{ulem}
\usepackage{varioref}

\begin{document}

\begin{center}
{{{\Large \bf 
The Effect of Anisotropic Extra Dimension in Cosmology
}
}\\[17mm]

Seyen Kouwn$^{1,2}$,~~Phillial Oh$^{3}$, and~~Chan-Gyung Park$^{4}$\\[3mm]

{\it$^{1}$Institute of Convergence Fundamental Studies \& School of Liberal Arts,
Seoul National University of Science and Technology, Seoul 139-743, Korea\\[2mm]
$^{2}$Korea Astronomy and Space Science Institute,
Daejeon 305-348, Republic of Korea\\[2mm]
$^{3}$Department of Physics,~BK21 Physics Research Division,
~Institute of Basic Science,\\
Sungkyunkwan University, Suwon 440-746, Korea\\[2mm]
$^{4}$Division of Science Education and Institute of Fusion Science,
Chonbuk National University, Jeonju 561-756, Korea}\\[2mm]
{\tt seyenkouwn@gmail.com,~ploh@skku.edu,~parkc@jbnu.ac.kr} }
\end{center}

\vspace{10mm}

\begin{abstract}
We consider five dimensional conformal gravity theory 
which describes an anisotropic extra dimension.
Reducing the theory to four dimensions yields Brans-Dicke theory
with a potential and a hidden parameter $\alpha$ which implements the anisotropy between the  four dimensional spacetime and  the extra dimension.
We find that a range of value of the parameter $\alpha$ can address the current dark energy density compared to the Planck energy density.
Constraining the parameter $\alpha$ and the other cosmological model parameters using the recent observational data
consisting of the Hubble parameters, type Ia supernovae, and baryon acoustic oscillations, 
together with the Planck or WMAP 9-year data of the cosmic microwave background radiation,
we find $\alpha>-2.05$ for Planck data and $\alpha>-2.09$ for WMAP 9-year data at 95\% confidence level.
We also obtained constraints on the rate of change of the effective Newtonian constant~($G_{\rm eff}$) at present 
and the variation of $G_{\rm eff}$ since the epoch of recombination to be consistent with observation.
\end{abstract}
\newpage

\tableofcontents

\newpage
\section{Introduction}

Nowadays, research on the higher dimensional gravity theories like Kaluza-Klein theory, string theory,  and brane world scenario constitutes one of the mainstream of theoretical particle physics. In such theories, it is usually taken for granted that the higher dimensional spacetime is isotropic. Even though the isotropic spacetime appeals more aesthetical from the viewpoint of symmetry like Lorentz symmetry and general covariance,
this has never been experimentally  verified. Therefore, it is a fundamental question to ask whether   higher dimensional spacetime has uniform physical properties in all directions \cite{Long:2002wn,Joyce:2014kja} and envisage the possibility that the extra dimensions might not  share the same property with the four dimensional spacetime we are living in.

Recently, an attempt to construct a  higher dimensional gravity theory in which the four dimensional spacetime and extra dimensions are not treated on an equal footing was made \cite{Moon:2017rox}. It is based on two compatible
symmetries of foliation preserving diffeomorphism and anisotropic conformal transformation. The anisotropy is first implemented in the higher dimensional metric by keeping the general covariance only for the four dimensional spacetime. This was achieved by adopting foliation preserving diffeomorphism in which the foliation is adapted along the extra dimensions. Then, it was  extended to
conformal gravity  with
introduction of  conformal scalar field.
 In  order to realize the anisotropic conformal invariance a real parameter $\alpha$ which measures the degree of anisotropy of conformal transformation between the spacetime and extra dimensional metrics was introduced.
In the zero mode effective
four dimensional action, it reduces to a  scalar-tensor theory coupled with nonlinear sigma model described by extra dimensional metrics. There are no restrictions on the value of $\alpha$ at the classical level. In this paper, we present a cosmological test of the scalar-tensor theory thus obtained in the case of five dimensional theory and check whether or not a specific value of $\alpha$ is preferred.

In general, the conformal invariance constrains the theory in a very tight form in a conformal gravity \cite{Moon:2009zq}, and contains at most one parameter,
that is the potential coefficient $\lambda,$ $V(\phi)=\frac{\lambda}{4}\phi^4.$
The Brans-Dicke theory contains more parameters \cite{Will:2014kxa}: one is $\omega$, which is the ratio between the nonminimally coupled  $\phi^2 R$ term and kinetic energy term for $\phi.$ Others are the  potential and its respective coefficients, if introduced.
It turns out that in the five dimensional anisotropic conformal gravity, the effective four dimensional scalar-tensor theory reduces to the Brans-Dicke theory with
a potential, in which the parameter $\omega$ and the power of the potential, $V(\phi)\sim \phi^{n}$, are determined in term of the parameter $\alpha$. Therefore, from the view point of Brans-Dicke theory, $\alpha$ is a hidden parameter and this is a consequence of anisotropic conformal invariance in higher dimensions.

In the gravitational theory with anisotropic conformal invariance, it is more convenient to work with a dimensionless scalar field in order to countercheck the arbitrary anisotropy factor $\alpha$. Recall that the kinetic coefficient $\omega$  of the Brans-Dicke theory can be allowed to be an arbitrary (positive definite) function of the scalar field, $\omega=\omega(\phi),$ which results in a general class of scalar-tensor theories with a dimensionless scalar field and they can be tested with  the solar  system  experiments\cite{Will:2014kxa}. In our case, the scalar field is also dimensionless. Nevertheless, $\omega$ is constrained to be a constant for the sake of the anisotropic conformal invariance, rendering the theory to be a Brans-Dicke type.

Another important point to be mentioned is that in our four dimensional Brans-Dicke theory, the origin of the Brans-Dicke scalar can be identified with the conformal scalar that is necessarily introduced for the purpose of conformal invariance.
It is well-known that in the isotropic case, the conformal or Weyl scalar field is a ghost field with a kinetic coefficient yielding a negative kinetic energy and they cannot become the Brans-Dicke scalar \cite{Moon:2009zq}. However, in the anisotropic case, the kinetic coefficient $\omega$ is determined as a  specific function of $\alpha$ and there exists a range of parameter $\alpha$ where $\omega(\alpha)$
becomes positive. We will check that the actual cosmological test prefers the range of parameter $\alpha$ with a positive value of $\omega.$

The paper is organized as follows: In Sec. 2, we give a formulation of the 5D gravity with anisotropic conformal invariance and perform dimensional reduction to obtain 4D Brans-Dicke theory.
We perform cosmological analysis and give numerical results for evolution equations.
In Sec. 3, comparisons with the recent cosmological data are made and the range of parameter $\alpha$ is 
constrained.
Sec. 4 contains conclusion and discussion.

\section{Model}
We start with a formulation of 5D anisotropic conformal gravity.
The  first part of this section is mostly redrawn from Ref. \cite{Moon:2017rox}
to make the paper self-contained. Let us first consider the
Arnowitt-Deser-Misner (ADM) decomposition of five dimensional metric:
\begin{align}
ds^2 = g_{\mu\nu}( dx^\mu +N^\mu dy)( dx^\nu +N^\nu dy) + N^2 dy^2.
\end{align}
Then, the five dimensional Einstein-Hilbert action with cosmological constant is expressed as
\begin{align}
S_{\rm{EH}}=\int \,dy
d^4xN\sqrt{-g}~M_*^3\left[\left(R-2\Lambda_5\right)-\{K_{\mu\nu}K^{\mu\nu} -
K^2\}\right],
\label{5dR}
\end{align}
where $M_*$ is the five dimensional gravitational constant, $R$ is the spacetime curvature, $\Lambda_5$ is the cosmological constant, and $K_{\mu\nu}$ is the extrinsic curvature tensor,
$K_{\mu\nu}=(\partial_y g_{\mu\nu} - \nabla_{\mu} N_{\nu} -
\nabla_{\nu} N_{\mu})/(2N)$.
The above action (\ref{5dR}) can be extended anisotropically
by breaking the five dimensional general covariance down to its foliation preserving
diffeomorphism symmetry given by
\begin{align}
x^{\mu}\to x^{\prime\mu}&\equiv x^{\prime\mu}(x,y),
~~~y\to y^{\prime}\equiv
y^{\prime }(y),\label{FPD}\\
g^{'}_{\mu\nu}(x',y')&=\left(\frac{\partial x^\rho}{\partial
x^{'\mu}}\right)\left(\frac{\partial
x^\sigma}{\partial x^{'\nu}}\right)g_{\rho\sigma}(x,y),\label{trans1}\\
{N'}_{}^{\mu}(x',y')&=\Big(\frac{\partial y}{\partial
y^{'}}\Big)\Big[\frac{\partial x^{'\mu}}{\partial
x^{\nu}}N_{}^{\nu}(x,y)-\frac{\partial x'^{\mu}}{\partial
y^{}}\Big],\label{trans2}\\
N^{\prime}(x',y')&=\left(\frac{\partial y}{\partial y{'}}\right)N(x,y)
,
\end{align}
and non-uniform conformal transformations
\begin{eqnarray}
g_{\mu\nu}\rightarrow e^{2\omega(x,y)}g_{\mu\nu},~~
N\rightarrow e^{\alpha \omega(x,y)}N,~~ N^{\mu}\rightarrow N^{\mu},
~~\varphi\rightarrow
e^{-\frac{\alpha+2}{2}\omega}\varphi,\label{confotrans11}
\end{eqnarray}
where a Weyl scalar field $\varphi$ to compensate the conformal transformation
of the metric is introduced.
In the above equation \eqref{confotrans11}, a factor $\alpha$ is introduced in the transformation of $N(=g_{55})$, 
which characterizes the anisotropy of spacetime and
extra dimension
\footnote{
We assume that the the  field $\varphi$  is a dimensionless and $M_*$
is a scale related with Planck scale.
We also consider only the case $\alpha \neq -2$, because $\varphi$ is not effected
under the conformal transformation in \eqref{confotrans11}.
 It can be actually shown that for  $\alpha = -2$, an anisotropic scale invariant gravity
theory can be constructed without the need of the field $\varphi$. 
}.
The anisotropic Weyl action invariant under Eqs. \eqref{FPD}-\eqref{confotrans11} for an arbitrary $\alpha$ can be written as
\begin{eqnarray}
&&\hspace*{-2em}S=\int dy
d^4x\sqrt{-g}NM_{*}^{3}\Bigg[\varphi^{2}\left(R
-\frac{12}{\alpha+2}\frac{\nabla_{\mu}\nabla^{\mu}\varphi}{\varphi}
+\frac{12\alpha}{(\alpha+2)^2}\frac{\nabla_{\mu}\varphi\nabla^{\mu}\varphi}{\varphi^2}
\right)\nonumber\\
&&\hspace*{7em}~
-\beta_1\varphi^{-\frac{2(\alpha-4)}{\alpha+2}}\left\{B_{\mu\nu}B^{\mu\nu} -
\lambda B^2\right\}+\beta_2 \varphi^{2}A_{\mu}A^{\mu}-V(\varphi)\Bigg].\label{conformalR}
\end{eqnarray}
where $\beta_1, \beta_2, \lambda$ are some constants,
the potential $V$, $B_{\mu\nu}$ and $A_\mu$ are given by
\begin{align}
V&=V_0\varphi^{\frac{2(\alpha+4)}{\alpha+2}},\label{fracpote} \\
B_{\mu\nu}&=K_{\mu\nu}+\frac{2}{(\alpha+2)N\varphi}g_{\mu\nu}(\partial_y
\varphi-\nabla_{\rho}\varphi N^{\rho})\,,\quad
B\equiv g^{\mu\nu}B_{\mu\nu},\\
A_{\mu}&=\frac{\partial_{\mu}N}{N}
+\frac{2\alpha}{\alpha+2}\frac{\partial_{\mu}\varphi}{\varphi}.
\end{align}

A couple of comments are in order. The isotropic case with $\beta_1=\lambda=\alpha=1,$ and  $\beta_2=0$ leads to five dimensional Weyl gravity with a potential $V\sim \phi^{\frac{10}{3}}$ \cite{Moon:2009zq}. In the anisotropic case, the action (\ref{conformalR})
is, in general, plagued with perturbative ghost instability coming from breaking of the full general covariance of 5D. However, it can be shown that this problem can be cured by constraining  the constants  $\beta_1$ and $\beta_2$, especially with $0<\beta_2<\frac{3}{2}$ \cite{Moon:2017rox}.

Now we discuss 4-dimensional effective low energy action and let us consider only
zero modes.
We first go to a ``comoving'' frame with $N^{\mu}=0$  and
 impose $y$-independence (cylindrical condition) for $g_{\mu\nu}=g_{\mu\nu}(x), \phi=\phi(x)$ and $N=N(x)$.
This enables to replace $\int dy=L$ where $L$ is the
size of the extra dimension and eliminates terms containing $B_{\mu\nu}$
and $B.$ The resulting action preserves the redundant conformal transformation
\begin{align}
g_{\mu\nu}\rightarrow e^{2\omega(x)}g_{\mu\nu},~~
N\rightarrow e^{\alpha\omega(x)}N,~~ \varphi\rightarrow
e^{-\frac{\alpha+2}{2}\omega(x)}\varphi,\label{confotrans2}
\end{align}
where $\omega(x,y)$ in (\ref{confotrans11}) is replaced with $\omega(x).$
 Using  this, we further fix $N(x)=1$ and
 find the resulting four dimensional action given by
\begin{align} \label{IGaction}
S &=  \int d^4 x \sqrt{-g}\left[
{\gamma_1 M_p^2\over 2} \varphi^2 R
-{\gamma_1 M_p^2 \omega\over 2} \nabla_\mu \varphi \nabla^\mu \varphi
- \gamma_2 M_p^4 \varphi^{ 2\alpha+8 \over \alpha+2}
\right] \,,
\end{align}
where $\gamma_1$ and $\gamma_2$ are defined as
\begin{align} \label{gammadefine}
M_*^3L\equiv \gamma_1M_p^2/2 \,, \quad
M_*^3LV_0\equiv \gamma_2M_p^4,
\end{align}
and $\omega$ is given by
\begin{align}\label{BDparam}
\omega \equiv {-4(\alpha+1)(\beta_2 \alpha+6)\over (\alpha+2)^2}
\,.
\end{align}
Let us redefine the field as
\begin{align*}
\varphi \rightarrow \tilde{\varphi} = \sqrt{\gamma_1} \varphi.
\end{align*}
Then, the action (\ref{IGaction}) becomes
\begin{align}
S &=  \int d^4 x \sqrt{-g}\left[
{M_p^2\over 2} \tilde{\varphi}^2 R
-{ M_p^2 \omega\over 2} \nabla_\mu \tilde{\varphi} \nabla^\mu \tilde{\varphi}
- \tilde{V}(\tilde\varphi)
\label{bdaction}
\right] \,,
\end{align}
where
\begin{align*}
\tilde{V}(\tilde\varphi) = \gamma_2 \gamma_1^{-{\alpha+4\over \alpha+2}}M_p^4\tilde{\varphi}^{ 2\alpha+8 \over \alpha+2} \equiv\tilde V_0\tilde{\varphi}^{ 2\alpha+8 \over \alpha+2}\,.\label{smallpot}
\end{align*}
We find the effective four dimensional action is given by
Brans-Dicke theory with a potential. $\omega$ is positive
for a range of $-\beta_2^{-1}6<\alpha<-1$. Note that $\varphi$ is usually  a ghost field
in the isotropic case with  $\omega=-16/3$. This status has been evaded for anistropic case, thus reproducing the Brans-Dicke theory in the effective four dimensional action.
 Moreover,  $\omega$ becomes a big number for $\alpha$
being close to $-2$, as is usually required to pass the
solar system test. We will show that
$\alpha$ close to $-2$ is indeed preferred in the cosmological test, which implies
$\gamma_1 \sim {\cal O}(0.1)$ and $\gamma_2 \sim {\cal O}(1)$.
 It corresponds to Kaluza-Klein reduction with $M_*^3\sim \gamma/
 {G_5}$, with $G_5$ being the five dimensional Newton's constant.

From here on, we remove the tilde notation in Eq. (\ref{bdaction}).
We include matter term to investigate the cosmology
\footnote{We assume that the matter term couples with only $g_{\mu\nu}$
and breaks the anisotropic conformal invariance from the beginning.}. The Einstein equations obtained from action \eqref{IGaction} by varying with respect to the metric
$g_{\mu\nu}$ can be written in the following form:
\begin{align}
\varphi^2 G_{\mu\nu} &= {1\over  M_p^2} T_{\mu\nu}^{(m)}
+2 \varphi \nabla_\mu \nabla_\nu \varphi
+\left(2+\omega \right)\nabla_\mu\varphi \nabla_\nu\varphi \nn \\
&\quad\quad
+g_{\mu\nu} \left[
-2 \varphi \Box \varphi
-\left(2+{\omega\over2}\right) \nabla_\alpha\varphi \nabla^\alpha \varphi
-V
\right]
\,,
\end{align}
and scalar field equation is given by
\begin{align}
 M_p^2 \omega\Box \varphi +  M_p^2 \varphi R - V_\varphi = 0\,.
\end{align}
In this work we shall study the isotropic and homogeneous cosmology.
Thus, we consider the  space-time geometry is given by the Robertson-Walker metric:
\begin{align}
ds^2 = -dt^2 + a^2\left(
{dr^2\over 1 -k r^2}+ r^2d\theta^2+r^2 \sin^2\theta d\phi^2 
\right)
\,.
\end{align}
In this metric, Einstein equations follow as
\begin{align}
3H^2 &= {\omega \over 2 }{\dot{\varphi}^2 \over \varphi^2}
-6H{\dot{\varphi} \over \varphi} +{V \over  M_p^2 \varphi^2}
+{\rho_{r,0} \over  M_p^2 \varphi^2 a^4}
+{\rho_{m,0} \over  M_p^2 \varphi^2 a^3} 
-{3k \over M_{\rm p}^2 \varphi^2 a^2}\,, \label{eq1} \\
-3H^2-2\dot{H} &= {2\ddot{\varphi} \over \varphi} + 4H{\dot{\varphi} \over \varphi}
+\left(2+{ \omega \over 2}\right){\dot{\varphi}^2 \over \varphi^2}
-{V \over  M_p^2 \varphi^2}
+{\rho_{r,0} \over 3  M_p^2 \varphi^2 a^4}
+{k \over M_{\rm p}^2 \varphi^2 a^2}  \label{eq2} \,,
\end{align}
where $H\equiv \dot{a}/a$ is the Hubble parameter and $\rho^{(r)}$ and $\rho^{(m)}$ are 
the standard radiation and matter
energy densities. The field equation for the scalar field can be rewritten as
\begin{align}
\ddot{\varphi} + 3H\dot{\varphi}-{6 \over \omega }(2H^2+\dot{H})\varphi
+{V_\varphi\over \omega  M_p^2 }
=0 \label{eq3} \,,
\end{align}
where $V_\varphi$ denotes the derivative of the potential $V(\varphi)$ with respect to $\varphi$.

As in~\cite{Boisseau:2000pr}, the effective Newtonian constant is
\begin{align}\label{Geff}
G_{\rm eff} = {G_{\rm N}\over \varphi^2 } {1+8 /\omega \over 1+6 /\omega} \,,
\end{align}
where $G_{\rm N} = 6.67 \times 10^{-8} {\rm cm}^3 {\rm g}^{-1} {\rm s}^{-2}$ is Newton's constant
measured in Cavendish-type and solar system experiments
\footnote{Note that when $\varphi$ is fixed to be a constant $\varphi=\varphi_*$, $G_{\rm eff*}$ becomes the Newton's coupling constant which corresponds to the vacuum solution of a de Sitter universe in the induced gravity context~\cite{Cooper:1982du}.}.
The present value of $\varphi_0$ can be connected the Newton's constant $G_{\rm N}$
by the relation:
\begin{align}\label{phi0}
\varphi_0^2 =
{1+8 /\omega \over 1+6 /\omega} \,.
\end{align}
In order to compare with dark energy in Einstein gravity with a Newton's constant $G_N$,
we identify dark energy density and pressure as follows~\cite{Finelli:2007wb,Umilta:2015cta,Ballardini:2016cvy}:
\begin{align}
\rho_{\rm DE}  &=
{\varphi_0^2 \over \varphi^2}\left(
{\omega  M_p^2 \over 2  }\dot{\varphi}^2
-6  M_p^2H\varphi\dot{\varphi} +V
\right)
+  \left({\varphi_0^2 \over \varphi^2}-1\right){\rho_{r,0} \over a^4}
+  \left({\varphi_0^2 \over \varphi^2}-1\right){\rho_{m,0} \over a^3}
-  \left({\varphi_0^2 \over \varphi^2}-1\right){3k \over a^2}
\,, \label{definedarkenergy} \\
p_{\rm DE} &=
{\varphi_0^2 \over \varphi^2}\left[
2 M_p^2 \varphi \ddot{\varphi}
+4  M_p^2H\varphi\dot{\varphi}
+ M_p^2\left(
2 + {\omega \over 2  }
\right)\dot{\varphi}^2
-V
\right]
+ {1\over 3} \left({\varphi_0^2 \over \varphi^2}-1\right){\rho_{r,0} \over a^4}
+ \left({\varphi_0^2 \over \varphi^2}-1\right){k \over a^2}
\,.
\end{align}
 
\begin{figure}[ht]
\begin{center}
\scalebox{0.9}[0.9]{
\includegraphics{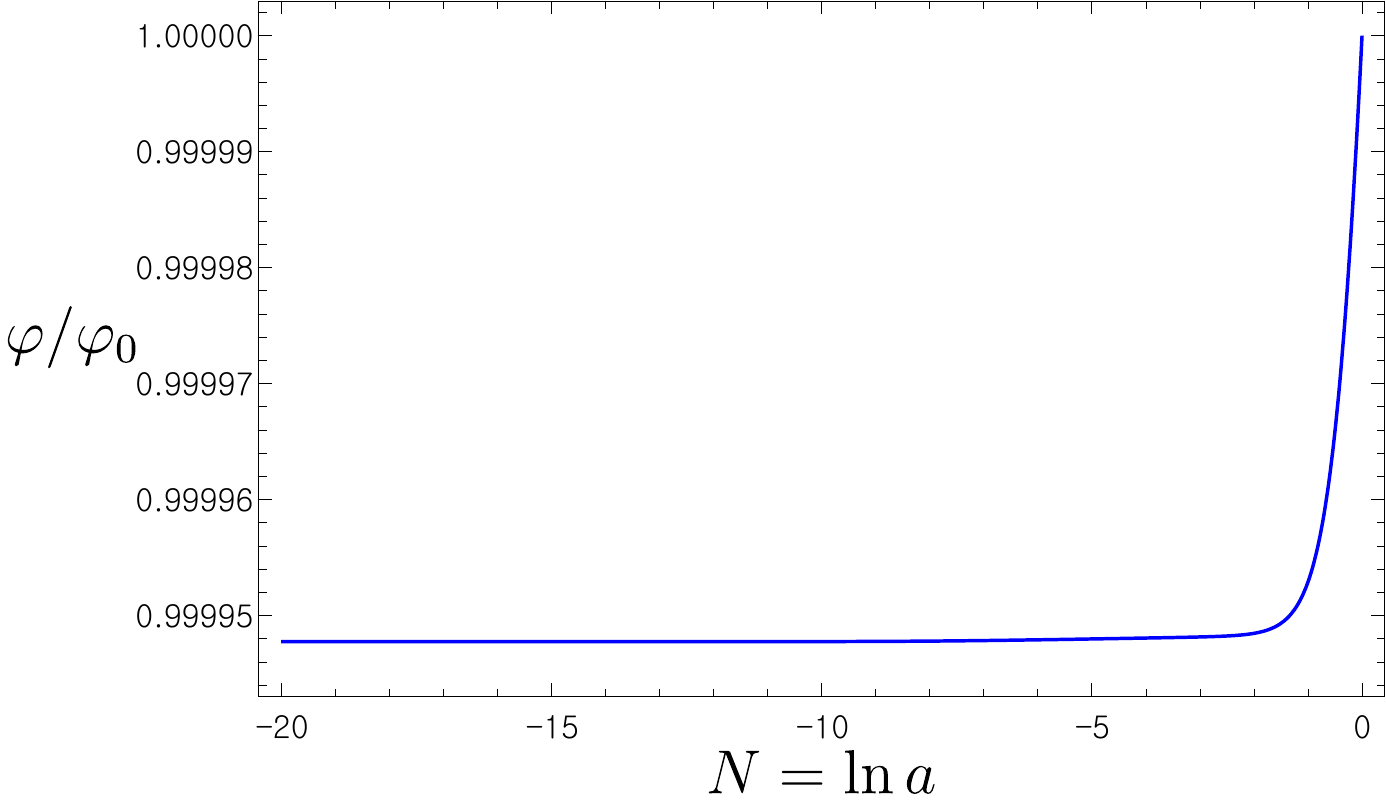}
}
\end{center}
\caption{\small
The typical evolution of the Brans-Dicke field $\varphi$ from radiation
domination epoch to the present time as function of $N=\ln a$.
Here, the cosmological parameters are $\alpha=-2.001$, $V_0=0.3$, $\beta_2=1$, $\Omega_r h^2=4.17\times 10^{-5}$
, $\Omega_k h^2=0$
and $\Omega_m h^2 = 0.14$.
}\label{fig:phisol}
\end{figure}

\begin{figure}[ht]
\begin{center}
\scalebox{0.55}[0.55]{
\includegraphics{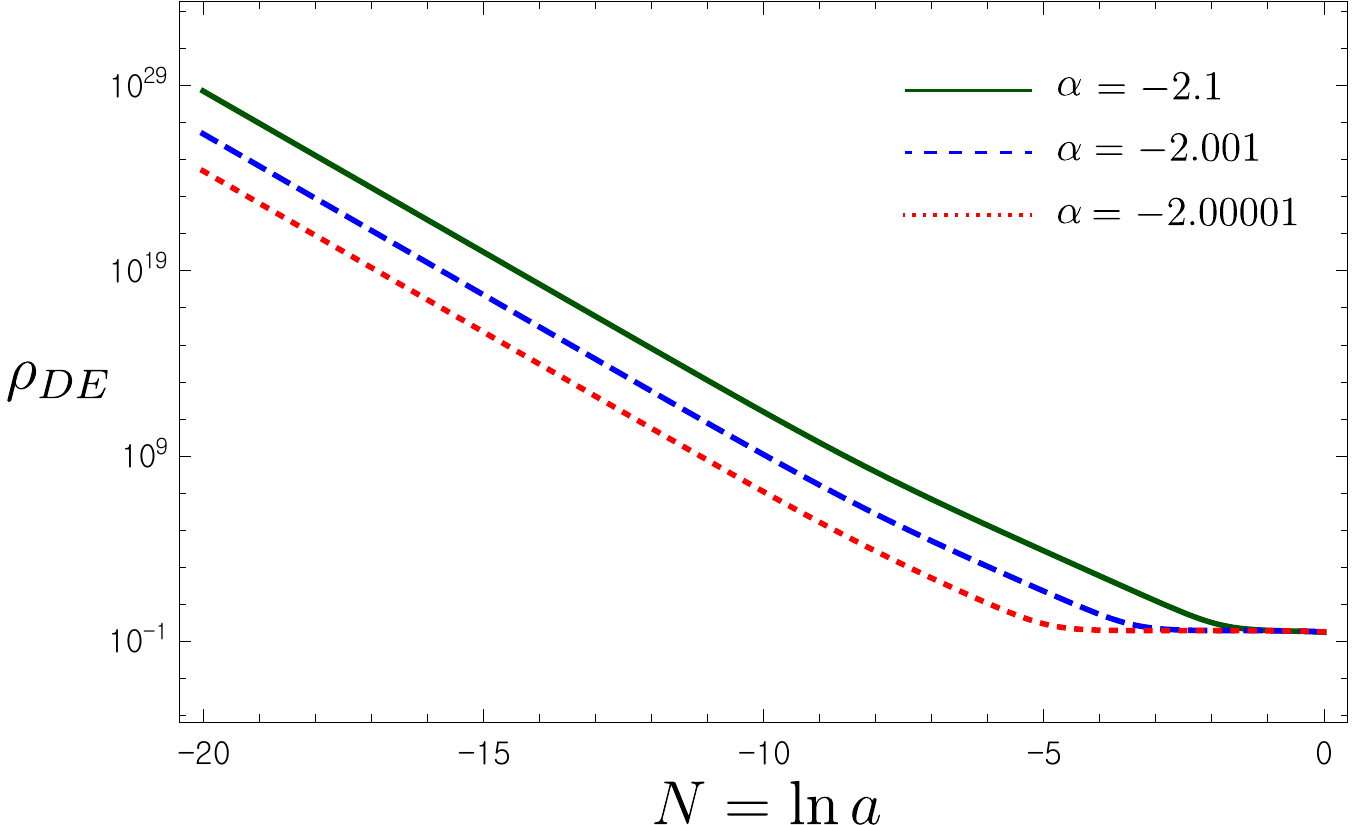}\quad
\includegraphics{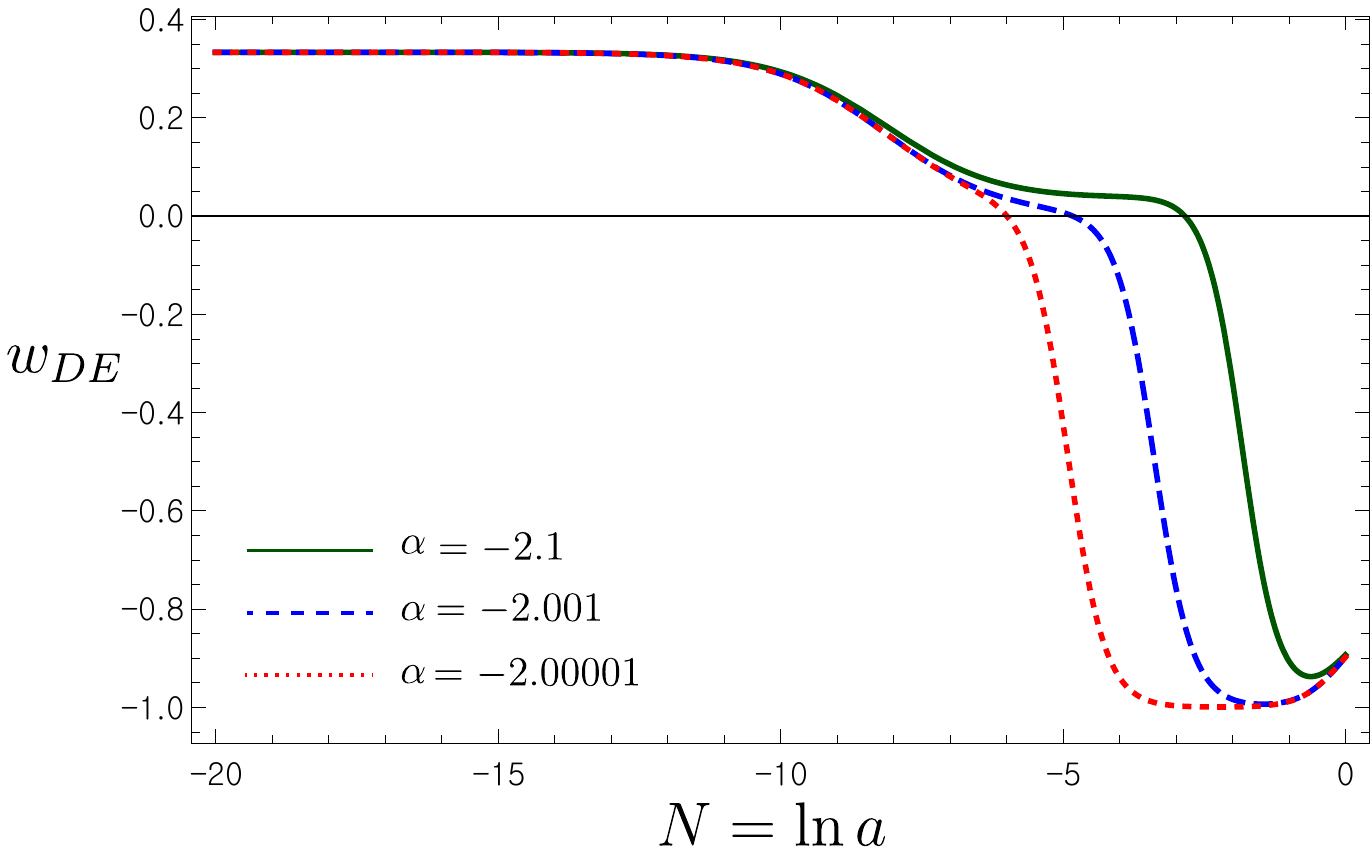}
}
\end{center}
\caption{\small
Evolution of energy density $\rho_{\rm DE}$ (left)
and equation of state parameter $w_{\rm DE}$ (right)
for different values of $\alpha$.
Here, the cosmological parameters are $V_0=0.3$, $\beta_2=1$, $\Omega_r h^2=4.17\times 10^{-5}$,
$\Omega_k h^2=0$ and $\Omega_m h^2 = 0.14$.
}
\label{fig:energyeos}
\end{figure}

Before comparing with observational data, we first perform a numerical analysis of 
the background evolutions equations.
Basically, the evolutions are determined by the parameters
$\alpha$, $V_0$, $\beta_2$ and the present matter density $\rho_{i,0}$.
The Brans-Dike parameter $\omega$ and the current value of scalar field $\varphi_0$
are determined by the relations \eqref{BDparam} and \eqref{phi0}, respectively.
For initial conditions of the scalar field $\varphi$,
the initial velocity $\dot{\varphi_i}$ is assumed to be zero,
and the initial value $\varphi_i$ is determined by the shooting method.
In Fig.~\ref{fig:phisol} we plot the evolution of the $\varphi$.
At early times the scalar field dynamcis is frozen during the radiation-dominated epoch
and begins to grow at the end of radiation-dominated epoch to realize the Newtonian gravitational constant
at present time. 
In the first of Fig.~\ref{fig:energyeos} we display the time evolution of the energy density 
for dark energy defined \eqref{definedarkenergy}.
It has the same scaling behavior with the radiation energy density at early time
and eventually it remains almost constant near the present time.
In the second of Fig.~\ref{fig:energyeos} we plot the time evolution of the equation of state parameter for dark energy $w_{DE} \equiv p_{DE}/\rho_{DE}$.
We again find that its value is close to $1/3$ in the early radiation dominant epoch
and approche $-1$ until recently.
It is worth mentioning that this behavior is similar in some respects to one in~\cite{Kamenshchik:2012rs} 
although their origins are different,
where the scalar field stays in a minimum of potential until radiation epoch 
and is shifted from the minimum during the transition from radiation to matter
which leads to a suitable amount of dark energy explaining the present accelerated expansion.

\newpage
\section{Observational Constraints}\label{secob}
In this section we constrain our model
with the latest cosmological data
described in~\cite{Kouwn:2015cdw},
and investigate whether or not
it can be distinguished from the $\Lambda$-CDM model.
For this purpose,
we use the recent observational data such as type Ia supernovae (SN),
baryon acoustic oscillation (BAO) imprinted in large-scale structure of galaxies,
cosmic microwave background radiation (CMB), and Hubble parameters [$H(z)$].
For numerical analysis,
it is convenient to rewrite equations~\eqref{eq1}-\eqref{eq2}
in terms of
$N\equiv \ln a$ as follows:
\begin{align}\label{eqH2com}
\hat{H}^2 = &{\omega\over 6} \hat{H}^2\left({\varphi'\over \varphi}\right)^2
-2\hat{H}^2\left({\varphi'\over \varphi}\right)
+\hat{V}_0 \varphi_0^2 \varphi^{4\over \alpha+2}
+\Omega_m h^2 \left({\varphi_0^2\over \varphi^2}\right) e^{-3N} \nn \\
&+\Omega_r h^2 \left({\varphi_0^2\over \varphi^2}\right) e^{-4N}
+\Omega_k h^2 \left({\varphi_0^2\over \varphi^2}\right) e^{-2N}
\,,
\end{align}
where a prime indicates a derivative with respect to $N$. 
The equation of motion for the scalar field is rewritten as
\begin{align}
\hat{H}^2 \varphi'' + \left(3\hat{H}^2 + \hat{H}\hat{H}'\right)\varphi'
-{6\over \omega }\left(2\hat{H}^2 + \hat{H}\hat{H}'\right)\varphi
+ {3\hat{V}_0 \over \omega}\left({2\alpha+8 \over \alpha+2}\right) \varphi_0^2 \varphi^{\alpha+6\over \alpha+2}
=0
\,,
\end{align}
where we have introduced dimensionless quantities,
\begin{align}\label{dimlessvar}
\hat{H}^2 &\equiv {H^2 h^2\over H_0^2} \,, \quad
\Omega_r \equiv {\rho_{r,0} \over 3 H_0^2  M_p^2 \varphi_0^2} \,, \quad
\Omega_m \equiv {\rho_{m,0} \over 3 H_0^2   M_p^2 \varphi_0^2} \,, \quad
\nonumber \\
\Omega_k &\equiv { {-k} \over H_0^2  M_{\rm p}^2 \varphi_0^2} \,, \quad
\varphi_0 \equiv   {1\over \gamma} \left( { 1+8/\omega \over 1+6/\omega } \right)  \,, \quad
\hat{V}_0 \equiv {V_0 h^2 \over 3 H_0^2  M_p^2 \varphi_0^2}  \,,
\end{align}
Here, $H_0$ is the present value of the Hubble parameter,
usually expressed as $H_0=100 \,h\, {\rm km}\,\,{\rm s}^{-1}{\rm Mpc}^{-1}$,
$\Omega_r$ and $\Omega_m$ are the radiation and matter density parameters at the present epoch, respectively.
The radiation density includes the contribution of relativistic neutrinos as well as that of photons,
with the collective density parameter
\begin{align}
\Omega_r h^2 = \Omega_\gamma h^2 \left(1+0.2271N_{\rm eff}\right)
\,,
\end{align}
where $N_{\rm eff}=3.04$ is the effective number of neutrino species, and
$\Omega_\gamma$ is the photon density parameter with $\Omega_\gamma=2.47037\times 10^{-5} h^{-2}$ for the present CMB temperature $T_0=2.725~{\rm K}$~(WMAP9) and $\Omega_\gamma=2.47218\times 10^{-5} h^{-2}$ for $T_0=2.7255~{\rm K}$~(PLANCK).
Notice that the background dynamics is completely determined by a set of parameters
$(\alpha, \hat{V}_0, \Omega_m, \Omega_k, \beta_2)$.
We need the baryon density parameter ($\Omega_b$) to confront our model with the BAO and CMB data, 
and finally our model has six free parameters
$\mbox{\boldmath $\theta$}=(\alpha, \hat{V}_0, \Omega_b h^2, \Omega_m h^2, \Omega_k h^2, \beta_2)$.
It should be emphasized that the Hubble constant ($H_0$) is no longer a free parameter
because it is derived from the integration of field equations for a given set of parameters chosen.
The free parameters take the following priors:
$\alpha = [-3, -2]$, $\hat{V}_0 = [1,7]$,
$\Omega_b h^2 = [0.015, 0.030]$, $\Omega_m h^2 = [0.11, 0.15]$, $\Omega_k h^2 = [-0.1, 0.1]$
and $\beta_2 = [0, 1.5]$.
We apply the Markov chain Monte Carlo (MCMC) method to obtain the likelihood distributions for the model parameters \cite{MCMC}.
The method propagates the parameter vector $\mbox{\boldmath $\theta$}$ in random directions to explore the parameter space that is favored by the observational data, by making decisions for accepting or rejecting
a randomly chosen parameter vector (or chain element)
via the probability function
$P(\mbox{\boldmath $\theta$}|\mathbf{D}) \propto \exp(-\chi^2/2)$,
where $\mathbf{D}$ denotes the data,
and $\chi^2= \chi_{H(z)}^2+\chi_\textrm{SN}^2 + \chi_\textrm{BAO}^2 + \chi_\textrm{CMB}^2
$ is the sum of individual chi-squares
for $H(z)$, SN, BAO, and CMB data (defined below).
We consider that the convergence of the MCMC chain is achieved 
if the means estimated from the first (after burning process)
and the last 10\% of the chain are approximately equal to each other.

\subsection{Hubble Parameters}
We use 29 data points of the Hubble parameters in a redshift range of
$0.07 \leq z \leq 2.34$, which include 23 data points obtained from the differential age approach~\cite{Hubble:DA} and 6 derived from the BAO measurements~\cite{Hubble:BAO}. The chi-square is defined as
\begin{align}
\chi^2_{H(z)} = \sum_{i=1}^{29}
\frac{ \left[H_{\textrm{th}}(z_i) - H_{\textrm{obs}}(z_i)  \right]^2}{\sigma^2_H(z_i)}
\,,
\end{align}
where $H_{\textrm{th}}(z_i)$ and $H_{\textrm{obs}}(z_i)$ are theory-predicted and observed
values of the Hubble parameter at redshift $z_i$, respectively, 
and $\sigma_H$ indicates the measurement uncertainty
of the observed data point.

\subsection{Type Ia Supernovae}
In our analysis, the Union 2.1 compilation of 580 SNe in a redshift range of $0.015 \leq z \leq 1.414$ is 
used to constrain the energy content of the late-time Universe \cite{Suzuki:2011hu}.
We use the chi-square that has been marginalized over the zero-point uncertainty
due to the absolute magnitude and Hubble constant \cite{SNIa-MRG}:
\begin{align}
    \chi_{\textrm{SN}}^2 = c_1 - c_2^2 / c_3,
\end{align}
where
\begin{align}
     c_1 = \sum_{i=1}^{580}
       \left[ \frac{\mu_{\rm th}(z_i)-\mu_\textrm{obs}(z_i)}{\sigma_i}\right]^2,
      \quad
     c_2= \sum_{i=1}^{580}
        \frac{\mu(z_i)_{\rm th}-\mu_\textrm{obs}(z_i)}{\sigma_i^2},
       \quad
     c_3= \sum_{i=1}^{580} \frac{1}{\sigma_i^2},
\end{align}
where $\mu_\textrm{obs}(z_i)$ and $\sigma_i$ denote the observed distance modulus and 
its measurement uncertainty of SN at redshift $z_i$. 
The theoretical prediction of the distance modulus $\mu_{\rm th}$ is defined as
\begin{align}
\mu_{\rm th}(z)=5\log \left[ { (1+z) r(z) \over 10 {\, \rm pc} } \right]\,,
\end{align}
where $r(z)$ is the comoving distance at redshift $z$,
\begin{align}
   r(z)=\frac{c}{H_0 \sqrt{\Omega_k}}
       S_k \left[\sqrt{\Omega_k}\int_0^z \frac{H_0}{H(z')} dz' \right],
\end{align}
with $c$ the speed of light and $S_k[x]=\sin x\,, x\,, \sinh x$
for $\Omega_k<0 \,, \Omega_k=0\,, \Omega_k>0$, respectively.

\subsection{Baryon Acoustic Oscillations}
As the BAO parameter, we use six numbers of $r_s(z_d)/D_V(z)$ extracted
from the Six-Degree-Field Galaxy Survey \cite{6dFGS},
the Sloan Digital Sky Survey Data Release 7 and 9 \cite{SDSS},
and the WiggleZ Dark Energy Survey \cite{WiggleZ}.
These BAO data points were used in the WMAP 9-year analysis \cite{WMAP9}.
Here $D_V(z)$ is the effective distance measure related to the BAO scale
\cite{Eisenstein-etal-2005},
\begin{align}
   D_V(z) \equiv \left[r^2 (z) \frac{cz}{H(z)} \right]^\frac{1}{3}\,,
\end{align}
and $r_s(z_d)$ is the comoving sound horizon size at the drag epoch.
We use a fitting formula for the redshift of drag epoch ($z_d$)
\cite{Eisenstein-Hu-1998}:
\begin{align}
   z_d = \frac{1291(\Omega_m h^2)^{0.251}}{1+0.659(\Omega_m h^2)^{0.828}}
       \left[1 + b_1 (\Omega_b h^2)^{b_2} \right],
\end{align}
where
\begin{align}
   b_1 =0.313 (\Omega_m h^2)^{-0.419}
       \left[1+0.607(\Omega_m h^2)^{0.674} \right], \quad
   b_2 =0.238 (\Omega_m h^2)^{0.223}.
\end{align}
Since the sound speed of baryon fluid coupled with photons ($\gamma$)
is given as
\begin{align}
   c_s^2 = \frac{\dot{p}}{\dot\rho}
         = \frac{\frac{1}{3}\dot\rho_\gamma}{\dot\rho_\gamma + \dot\rho_b}
         = \frac{1}{3\left[1+(3\Omega_b/4\Omega_\gamma)a\right]},
\end{align}
the comoving sound horizon size before the last scattering becomes
\begin{align}
   r_s(z) = \int_0^t c_s dt'/a
          = \frac{1}{\sqrt{3}} \int_0^{1/(1+z)}
               \frac{da}{a^2 H(a)[1+(3\Omega_b/4\Omega_\gamma)a]^\frac{1}{2}} \,.
\end{align}
The BAO measurements provide the following distance ratios~\cite{WMAP9}
\begin{align}
&\left<r_s(z_d)/D_V(0.1)\right> =0.336\,,\quad~~\,
\left<D_V(0.35)/r_s(z_d)\right> =8.88\,,\\
&\left<D_V(0.57)/r_s(z_d)\right> =13.67\,, \quad~
\left<r_s(z_d)/D_V(0.44)\right> =0.0916\,,\\
&\left<r_s(z_d)/D_V(0.60)\right> =0.0726\,, \quad
\left<r_s(z_d)/D_V(0.73)\right> =0.0592 \,,
\end{align}
together with the inverse of the covariance matrix between measurement uncertainties
\begin{align}
    \mathbf{C}_{\textrm{BAO}}^{-1} =
          \left( \begin{array}{cccccc}
          4444.4 & 0 & 0 & 0 & 0 & 0  \\
             0   & 34.602 & 0 & 0 & 0 & 0  \\
             0   & 0 & 20.661157 & 0 & 0 & 0  \\
             0   & 0 & 0 & 24532.1 & -25137.7 & 12099.1  \\
             0   & 0 & 0 & -25137.7 & 134598.4 & -64783.9  \\
             0   & 0 & 0 & 12099.1 & -64783.9 & 128837.6 \end{array} \right).
\end{align}
The chi-square is given as
\begin{align}
\chi_{\textrm{BAO}}^2
=\mathbf{X}^T
\mathbf{C}_{\textrm{BAO}}^{-1} \mathbf{X}
\,,
\end{align}
where
\begin{align}
\mathbf{X}=
\left( \begin{array}{c}
             r_s(z_d)/D_V(0.1) - 0.336\\
             D_V(0.35)/r_s(z_d)  - 8.88\\
             D_V(0.57)/r_s(z_d) - 13.67\\
             r_s(z_d)/D_V(0.44) - 0.0916\\
             r_s(z_d)/D_V(0.60) - 0.0726\\
             r_s(z_d)/D_V(0.73) - 0.0592 \end{array} \right)
\,.
\end{align}

\subsection{Cosmic Microwave Background Radiation}
We use the CMB distance priors based on WMAP 9-year data~\cite{WMAP9}
and Planck data~\cite{Ade:2013zuv} to constrain our model.
The first distance measure is the acoustic scale $l_A$ defined as
\begin{align}
   l_A = \pi \frac{r(z_*)}{r_s (z_*)}.
\end{align}
The decoupling epoch $z_*$ can be calculated from the fitting function
\cite{Hu-Sugiyama-1996}:
\begin{align}
   z_*=1048 [1+0.00124(\Omega_b h^2)^{-0.738}]
           [1+g_1(\Omega_m h^2)^{g_2}]\,,
\end{align}
where
\begin{align}
   g_1 = \frac{0.0783(\Omega_b h^2)^{-0.238}}{1+39.5(\Omega_b h^2)^{0.763}},
   \quad
   g_2 = \frac{0.560}{1+21.1(\Omega_b h^2)^{1.81}}.
\end{align}
The second distance measure is the shift parameter $R$ which is given by
\begin{align}
   R(z_*) = \frac{\sqrt{\Omega_m H_0^2}}{c} r(z_*).
\end{align}
Recently, Shafer \& Huterer \cite{Shafer:2013pxa} derived the distance priors $(l_a,R,z_{*})$ for the WMAP and Planck data as an efficient summary of CMB information. Hereafter, we use these priors to constrain our model parameters.

The CMB distance prior has been widely used to constrain the dark energy property. However, it is worth mentioning that using the CMB distance prior to constrain the model parameters has some limitation in that the estimate of the distance prior itself is model-dependent. As mentioned in \cite{Komatsu:2010fb}, the distance prior can be safely applied to constrain the dark energy model only when the model considered is based on the standard Robertson-Walker universe with the conventional radiation, matter, and neutrinos and on nearly power-law primordial power spectrum of curvature perturbations with negligible tensor modes. Since our model is based on the Robertson-Walker space-time geometry and the consequent background evolution equations include the matter and radiation together with the effective dark energy component, it is justified to use the WMAP and Planck distance priors to constrain our dark energy model at least in the background level.

\subsubsection{WMAP 9-year data}
According to WMAP 9-year observations (WMAP9) \cite{WMAP9},
the mean values for the three parameters
$(l_A, R, z_*)$ are \cite{Shafer:2013pxa}
\begin{align}
\left<l_A (z_*)\right> =301.98\,,\quad
\left<R(z_*)\right> =1.7302\,, \quad
\left<z_* \right> =1089.09\,.
\end{align}
The inverse of the covariance matrix between the parameter uncertainties is
\begin{align}
     \mathbf{C}_{\textrm{WMAP9}}^{-1} =
\left(
\begin{array}{ccc}
 3.13365 & 15.1332 & -1.43915 \\
 15.1332 & 13343.7 & -223.16 \\
 -1.43915 & -223.16 & 5.44598 \\
\end{array}
\right).
\end{align}
The chi-square is given as
\begin{align}
\chi_{\textrm{WMAP9}}^2
=\mathbf{X}^T
\mathbf{C}_{\textrm{WMAP9}}^{-1} \mathbf{X}
\,,
\end{align}
where
\begin{align}
    \mathbf{X}=
       \left( \begin{array}{c}
             l_A (z_*) - 301.98 \\
             R(z_*) - 1.7302 \\
             z_*    - 1089.09 \end{array} \right)
             \,.
\end{align}

\subsubsection{Planck data}

According to Planck observations (PLANCK) \cite{Ade:2013zuv},
the mean values for the distance priors
$(l_A, R, z_*)$ are given as \cite{Shafer:2013pxa}
\begin{align}
\left<l_A (z_*)\right> =301.65\,,\quad
\left<R(z_*)\right> =1.7499\,, \quad
\left< z_* \right> =1090.41\,.
\end{align}
Their inverse covariance matrix is
\begin{align}
 \mathbf{C}_{\textrm{Planck}}^{-1} =
\left(
\begin{array}{ccc}
 42.7223 & -419.678 & -0.765895 \\
 -419.678 & 57394.2 & -762.352 \\
 -0.765895 & -762.352 & 14.6999 \\
\end{array}
\right).
\end{align}
The chi-square becomes
\begin{align}
\chi_{\textrm{Planck}}^2
=\mathbf{X}^T
\mathbf{C}_{\textrm{Planck}}^{-1} \mathbf{X}
\,,
\end{align}
where
\begin{align}
    \mathbf{X}=
       \left( \begin{array}{c}
             l_A (z_*) - 301.65 \\
             R(z_*) - 1.7499 \\
             z_*    - 1090.41 \end{array} \right)
             \,.
\end{align}


\subsection{Results}

\begin{figure}[!ht]
\begin{center}
\scalebox{1}[1]{
\includegraphics{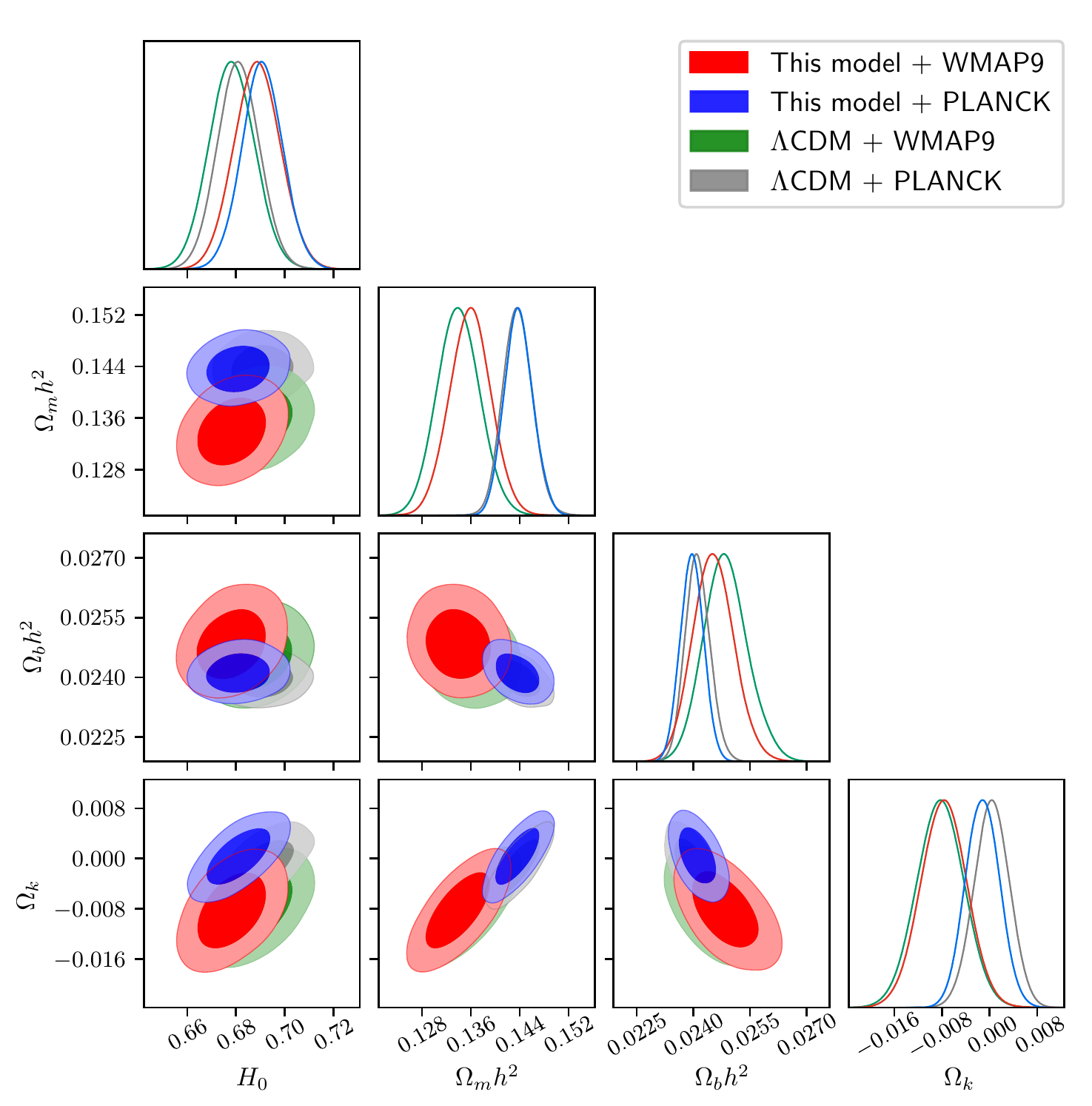}}
\end{center}

\caption{\small
Marginalized likelihood distributions of parameters that are common to our model and the $\Lambda$CDM model for different combinations of the data sets.
Here WMAP9 and PLANCK refer to $H(z)$ + SN+BAO+WMAP9 and $H(z)$ + SN+BAO+PLANCK, respectively.
The contours indicate 68\% and 95\% confidence limits.
}
\label{fig:like}
\end{figure}

\begin{figure}[!ht]
\begin{center}
\scalebox{0.8}[0.8]{
\includegraphics{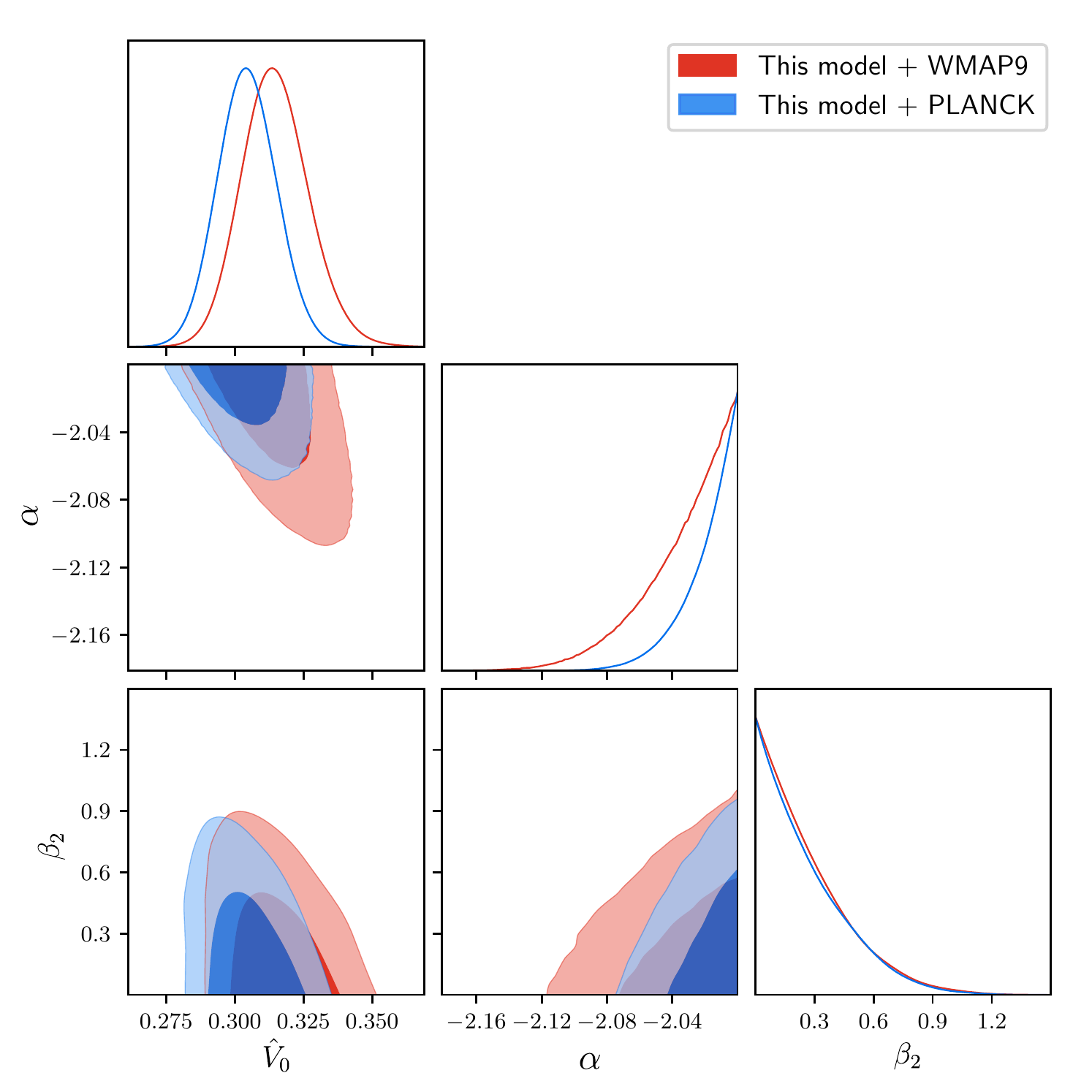}}
\end{center}
\caption{\small
Marginalized likelihood distributions of our model parameters $(\hat{V}_0, \alpha, \beta_2)$ with 68\% and 95\% confidence limits, obtained by the joint parameter estimation with $H(z)$ + SN+BAO+PLANCK (blue) and $H(z)$+SN+BAO+WMAP9 (red) data sets, respectively.
}
\label{fig:like2d}
\end{figure}

We explore the allowed ranges of our dark energy model parameters using the recent observational data by applying the MCMC parameter estimation method. In the calculation, 
we use $\alpha$, $\hat{V}_0$, $\Omega_m h^2$, $\Omega_b h^2$, $\Omega_k h^2$ and $\beta_2$ as free parameters. The results are shown in Table \ref{table:result1}
for a summary of parameter constraints with mean and $68\%$ confidence limits,
and Fig.~\ref{fig:like} for marginalized likelihood distributions of parameters that are common to our model and $\Lambda$CDM model.
The results for the other parameters of our model are presented in Fig.~\ref{fig:like2d}.
We can see that the result obtained with Planck data
gives tighter constraints on model parameters.
The best-fit locations in the parameter space are
\begin{align}
        (\alpha, \hat{V}_0, \Omega_m h^2, \Omega_b h^2, \Omega_k h^2, \beta_2)
=(-2.00052,  0.31295, 0.13330, 0.02469, -0.00382, 0.01017) \,,
\end{align}
with a minimum chi-square of $\chi_\textrm{min}^2 = 589.886$ for the $H(z)$+SN+BAO+WMAP9,
and
\begin{align}
(\alpha, \hat{V}_0, \Omega_m h^2, \Omega_b h^2, \Omega_k h^2, \beta_2)
=(-2.00089, 0.30259, 0.14328, 0.239700, 0.00002, 0.01282) \,,
\end{align}
with $\chi_\textrm{min}^2 = 599.747$ for $H(z)$+SN+BAO+PLANCK.


To assess the goodness-of-fit of our model, in Table \ref{table:result1}
we present the parameter constraints for the $\Lambda\textrm{CDM}$ model
and list the value of the minimum reduced chi-square ($\chi_\nu^2$) for each case.
The minimum reduced chi-square is defined as $\chi_\nu^2=\chi_{\textrm{min}}^2 /\nu$,
where $\nu=N-n-1$ is the number of degrees of freedom and $N$ and $n$ are the numbers of data points
and free model parameters, respectively.
In our analysis, $N=621$, and $n=6$ for our model
and $n=3$ for the $\Lambda\textrm{CDM}$ model.
Although the simple $\Lambda\textrm{CDM}$ model gives the slightly better fit to the observational data
with the smaller values of $\chi_\textrm{min}^2$ and $\chi_\nu^2$,
we judge that our model fits the data reasonably well
in the sense that the reduced chi-square is very close to unity.
We note that for our model to be compatible with observations
the parameter $\alpha$ must be lager than $\sim -2.1$
and the parameter $\hat{V_0}$ should be close to $0.3$,
which give the  following relation via \eqref{gammadefine} and \eqref{dimlessvar}.
\begin{align}
\gamma_2 \gamma_1^{-{\alpha+4 \over \alpha+2}} \sim 10^{-120}\,.
\end{align}
The above relation can be satisfied, for example,    with $\gamma_1 \sim {\cal O}(0.1)$ and $\gamma_2 \sim {\cal O}(1)$, in which case $\alpha \sim  -2.02$.

%




\begin{table*}
\begin{center}
\caption{Summary of parameter constraints and derived parameters.
The confidence levels are 68\% unless otherwise stated.}
\label{table:result1}
\resizebox{\columnwidth}{!}{%
\begin{tabular}{c||cc|cc}
\hline\hline
     &   \multicolumn{2}{|c|}{5D Brans-Dicke Model}    &   \multicolumn{2}{|c}{$\Lambda{\rm CDM}$ Model}   \\
\hline
 & $H(z)$ + SN + BAO &  $H(z)$ + SN + BAO & $H(z)$ + SN + BAO & $H(z)$ + SN + BAO \\
 & + WMAP9 & + PLANCK & +WMAP9 & +PLANCK\\
\hline

$H_0$ & $67.83^{+0.93}_{-0.93}$ & $68.10^{+0.86}_{-0.85}$
      & $68.87^{+0.94}_{-0.94}$ & $69.08^{+0.83}_{-0.82}$\\

$\Omega_m h^2$ & $0.1340^{+0.0035}_{-0.0035}$ & $0.1437^{+0.0024}_{-0.0024}$
               & $0.1359^{+0.0033}_{-0.0034}$ & $0.1438^{+0.0022}_{-0.0024}$ \\

$\Omega_b h^2$  & $0.02486^{+0.00054}_{-0.00062}$ & $0.02411^{+0.00031}_{-0.00034}$
                & $0.02453^{+0.00054}_{-0.00054}$ & $0.02397^{+0.00030}_{-0.00031}$\\

$\Omega_k $  & $-0.0083^{+0.0040}_{-0.0040}$ & $0.0004^{+0.0030}_{-0.0029}$
                & $-0.0077^{+0.0038}_{-0.0038}$ & $-0.0012^{+0.0028}_{-0.0028}$\\

$\beta_2$  & $< 0.699~(95\%~{\rm CL})$  & $< 0.680~(95\%~{\rm CL})$
                & - & - \\

$\alpha$ & $> -2.09~(95\%~{\rm CL})$  & $> -2.05~(95\%~{\rm CL})$
    & -                        & -\\

$\hat{V}_0$ & $0.315^{+0.011}_{-0.013}$ & $0.305^{+0.011}_{-0.011}$
            & -                            & -\\

$\Omega_\Lambda h^2$ & -                            & -
                     & $0.342^{+0.012}_{-0.012}$ & $0.334^{+0.011}_{-0.011}$ \\

\hline
$\chi^{2}_\textrm{min}$  & 589.886 & 599.747
                         & 584.344 & 590.502 \\
$\chi^{2}_\nu$  & 0.96073 & 0.97679
                & 0.94861 & 0.95861 \\
\hline
\hline
$|\gamma^{\rm PPN}-1|$ & $< 1.2\times10^{-3}~(95\%~{\rm CL})$  & $< 4.8\times 10^{-4}~(95\%~{\rm CL})$
                       & -                                & -\\

$\delta G /G$ & $  < 1.9\times 10^{-2}~(95\%~{\rm CL})$  & $< 9.5\times 10^{-3}~(95\%~{\rm CL})$
              & -                      & -\\

$\dot{G}/G~[ 10^{-13} {\rm yr}^{-1}]$ & $> -1.31~(95\%~{\rm CL})$  & $> -0.77~(95\%~{\rm CL})$
                                      & -                    & -\\
\hline
\hline
\end{tabular}
}
\end{center}
\end{table*}



\subsection{Local Constraints}
The general relativity in weak-field conditions
are confirmed by Solar-System experiments at the 0.04\% level~\cite{Williams:1995nq}.
Thus, we should verify that
the Brans-Dicke models
are presently close enough to Einstein's theory.
In the first post-Newtonian approximation of general relativity,
the deviations from general relativity can be parametrized by two real
numbers,
$\gamma^{\rm PPN}-1$ and $\beta^{\rm PPN}-1$, denoted by Eddington~\cite{Williams:1995nq}.
In the present models, they take the form
\begin{align}\label{ppn1}
\gamma^{\rm PPN}-1 = -{4 \over 8 + \hat{\omega}} \,, \quad
\beta^{\rm PPN} -1 = 0\,.
\end{align}
The Solar System experiments implies the following bounds
\begin{align}
& \left|2\gamma^{\rm PPN}-\beta^{\rm PPN}-1\right| < 3\times 10^{-3} \,, \\
& 4 \beta^{\rm PPN} - \gamma^{\rm PPN} -3 = -\left(0.7\pm 1\right)\times 10^{-3} \,, \\
& \left|\gamma^{\rm PPN}-1\right| = 4\times 10^{-4}\,, \\
& \gamma^{\rm PPN}-1 = \left(2.1\pm 2.3\right) \times 10^{-5}\,,
\end{align}
where the first bound was obtained from the perihelion shift of Mercury~\cite{perihelionshift},
the second from the Lunar Laser Ranging~\cite{Williams:1995nq},
the third from the light deflection observed by Very Long Baseline Interferometry~\cite{Shapiro:2004zz}
and the fourth from the Cassini mission~\cite{Bertotti:2003rm}.
These bounds can be resumed into the two limits~\cite{Schimd:2004nq}:
\begin{align}
& \left|\gamma^{\rm PPN}-1\right| < 2\times 10^{-3} \,, \quad
\left|\beta^{\rm PPN} -1\right| \leq 6\times 10^{-4} \,.
\end{align}
As a derived parameter, we quote the corresponding constraint
on the post-Newtonian parameter $\gamma^{\rm PPN}$ is
\begin{align}
&|\gamma^{\rm PPN}-1| < 1.2\times 10^{-3} \,,\quad (95\%~{\rm CL},~H(z)+{\rm SN+BAO+WMAP9}) \,,\\
&|\gamma^{\rm PPN}-1| < 4.8\times 10^{-4} \,,\quad (95\%~{\rm CL},~H(z)+{\rm SN+BAO+PLANCK}) \,.
\end{align}
In our models, the effective Newtonian constant~\eqref{Geff}
can vary from the recombination to the present epoch.
In order to place a constraint on the variation of the effective Newtonian constant,
we introduce two derived variables, namely,
the rate of change of the effective Newtonian constant $\dot{G}_{\rm eff}/G_{\rm eff}$ 
at present and the variation of effective Newtonian constant
$\delta G_{\rm eff}/G_{\rm eff}$ since the recombination epoch.
The theoretical expression for these variables are
\begin{align}\label{ppn2}
{\dot{G}_{\rm eff}\over G_{\rm eff}} = {-2 \dot{\varphi}_0 \over \varphi_0 } \,, \quad
{\delta {G}_{\rm eff}\over G_{\rm eff}} = {\varphi_{\rm rec}^{-2}-\varphi_0^{-2} \over \varphi_0^{-2} } \,.
\end{align}
Some previous constraints on $\dot{G}_{\rm eff} / G_{\rm eff}$ are summarized in Table~\ref{table:previousresult}.
We derive the following constraints on
the rate of change of the effective Newtonian constant at the present epoch,
\begin{align}
\dot{G}_{\rm eff} / G_{\rm eff} & > -1.31\times 10^{-13} ~{\rm yr}^{-1}\,,
\quad (95\%~{\rm CL},~H(z)+{\rm SN+BAO+WMAP9}) \,, \\
\dot{G}_{\rm eff} / G_{\rm eff} &> -0.77\times 10^{-13} ~{\rm yr}^{-1}\,,
\quad (95\%~{\rm CL},~H(z)+{\rm SN+BAO+PLANCK}) \,,
\end{align}
and on the variation of the effective Newton's constant
between the recombination and the present epochs:
\begin{align}
\delta {G}_{\rm eff} / G_{\rm eff} &< 1.9\times 10^{-2}\,,
\quad (95\%~{\rm CL},~H(z)+{\rm SN+BAO+WMAP9}) \,,\\
\delta {G}_{\rm eff} / G_{\rm eff}  &< 9.5\times 10^{-3}\,,
\quad (95\%~{\rm CL},~H(z)+{\rm SN+BAO+PLANCK}) \,.
\end{align}
Note that the constraints derived here are tighter than the previous constraints.
{\small
\begin{table*}
\begin{center}
\caption{Summary of constraints on the rate of change of the gravitational constant
$\dot{G}_{\rm eff}/G_{\rm eff}$. }
\label{table:previousresult}
\resizebox{\columnwidth}{!}{%
\begin{tabular}{|c|c|c|c|}
\hline
Author (year) & Phsical phenomena investigated & $\dot{G}_{\rm eff}/G_{\rm eff}~ [10^{-13} {\rm yr}^{-1}]$ & Ref.\\
\hline
Muller \& Biskupek (2007) & Lunar laser ranging     & $2\pm7$                      & \cite{Muller:2007zzb}\\
Copi (2004)\&Bambi (2005) & Big bang nucleosynthesis & $0\pm 4$                    &\cite{Copi:2003xd,Bambi:2005fi}\\
Guenther (1998)           & Helioseismology         & $0\pm 16$                    & \cite{Guenther:1998ApJ}\\
Thorsett (1996)           & Neutron star mass        & $-6\pm 20$                   & \cite{Thorsett:1996fr}\\
Hellings (1983)           & Viking lander ranging   & $20\pm 40$                   & \cite{Hellings:1983PRL} \\
Kaspi (1994)              & Binary pulsar            & $40\pm50$                    & \cite{Kaspi:1994hp} \\
Chang \& Chu (2007)       & CMB (WMAP3)             & $-96\sim 81~(95\%~{\rm CL})$     & \cite{Chan:2007fe} \\
Wu \& Chen (2010)         & CMB + LSS               & $-17.5 \sim 10.5~(95\%~{\rm CL})$ & \cite{Wu:2009zb} \\
Li et al. (2013)          & PLANCK + WP + BAO       & $-1.42^{+2.48}_{-2.27}$      & \cite{Li:2013nwa} \\
Li et al. (2015)          & PLANCK + BAO + SN       & $-2.65^{+1.83}_{-0.97}$      & \cite{Li:2015aug} \\
This paper                & PLANCK + BAO + SN+ $H(z)$ & $-1.31 \sim 0~(95\%~{\rm CL})$     & - \\
This paper                & WMAP9 + BAO + SN + $H(z)$ & $-0.77 \sim 0~(95\%~{\rm CL})$     & - \\
\hline
\end{tabular}
}
\end{center}
\end{table*}
}


\section{Discussion and Conclusion}\label{seccon}
In this paper, we
introduced a  five dimensional conformal gravity theory with anisotropic extra dimension
which is implemented by a  parameter $\alpha$.
Reducing the theory to four dimension yields Brans-Dicke theory
with  a potential;  $\omega $ and the potential are all determined in terms of the parameter $\alpha$ which is a hidden parameter from the four dimensional perspective.
For being compatible with the Solar System experiments
the Brans-Dicke parameter $\omega$ should be greater than 4000
which corresponds to  $\alpha$ being  close to $-2$.
Considering the case of Kaluza-Klein reduction,
$\gamma_1 \sim {\cal O}(0.1)$ and $\gamma_2 \sim {\cal O}(1)$,
the overall  potential energy density become  $\sim10^{-120} M_{\rm p}^4$
which sets the overall energy scale to be the current energy density.
From this point of view, the anisotropy of extra dimension
might be thought as being responsible for the extreme smallness of dark enrgy density.
Even though fine tuning is still required for $\alpha$,
it is fairly mild with $\sim {\cal O}(10^{-2})$ which can be compared to $\sim {\cal O}(10^{-120})$.
By applying the MCMC parameter estimation method,
we investigate the cosmological constraints on our model.
We found that the 95\% probability intervals for $\alpha$ parameter are
$\alpha>-2.05$ for PLANCK and $\alpha>-2.09$ for WMAP9,
which  corresponds to $\omega > 10300$ and $\omega > 4640$, respectivley.
We also derived the parametrized post-Newtonian parameters, and placed
the tightest cosmological constraints
on the corresponding derived post-Newtonian parameters.


The  extreme smallness of the  cosmological constant
with a negative $\alpha$ being close to $-2$ 
can  be addressed  in a different scheme.
Let us go back to the steps taken after Eq. (\ref{confotrans2}).
If instead of fixing $N(x)$=1, we perform conformal gauge fixing of  $\varphi(x)=\varphi_0,$
we obtain  from  (\ref{conformalR})
\begin{align}
S=\int
d^4x\sqrt{-g}\Bigg[\bar N\Bigl(R
-2\Lambda\Bigr)+\beta_2 \frac{g^{\mu\nu}\partial_{\mu}\bar N\partial_{\nu} \bar N}
{\bar N}\Bigg],
\label{fact}
\end{align}
with
\begin{align}
\bar N=\frac{ \gamma_1 \varphi_0^2 M_{pl}^2}{2}N, ~~\Lambda=\gamma_1^{-1}
\gamma_2\varphi_0^{\frac{4}{\alpha+2}}.\label{ccs}
\end{align}
We find that the reduced gravity corresponds to
Brans-Dicke theory
with a cosmological constant.
Note that $\Lambda$ in (\ref{ccs}) can yield a very small number for $\alpha$ being close to $-2. $
For example, with $\varphi_0\sim 10$ and $\alpha=-2.04$, the $\varphi_0$ part can produce a  number like $\sim 10^{-100}$. The more elaborate fine-tuning is necessary
in order to produce the correct factor for the cosmological constant problem, but fine-tuning problem can be substantially alleviated  compared to the conventional
one which requires a fine-tuning of the order
$10^{-120}$ for the cosmological constant.

It seems that the Brans-Dicke field $\bar N$ with
$0<\beta_2<\frac{3}{2}$ looks like a ghost field. But
if a further conformal transformation of metric $g_{\mu\nu}\rightarrow \chi^{-2}g_{\mu\nu}$
with $\chi^2=2\bar N/M_{pl}^2$ is ensued,
the action (\ref{fact}) becomes
$(Q\equiv M_{pl}\sqrt{\frac{3}{2}-\beta_2}\log(2\bar N/M_{pl}^2))$
\begin{align}
S=\int
d^4x\sqrt{-g}\Bigg[\frac{M_{pl}^{2}}{2}R
-\frac{1}{2} g^{\mu\nu}\partial_{\mu}Q
\partial_{\nu} Q-V(Q)\Bigg],
\label{fact1}
\end{align}
where the potential is given by
\begin{align}
V=V_0\ e^{-
\lambda Q/M_{pl}},  ~~ V_0=\Lambda M_{pl}^2,~~
\lambda=\left(\frac{3}{2}-\beta_2\right)^{-1/2}.
\end{align}
We find that the theory reduces to the exponential quintessence model.
The point is that the potential coefficient $V_0$ is proportional to the cosmological constant $\Lambda,$ which  can set the overall scale of the potential to be of the order of the present energy density, if $\alpha$
is suitably adjusted to be  close to $-2$ in (\ref{ccs}).  In this sense, it provides a chance to address the coincidence problem without extreme fine-tuning\cite{LopesFranca:2002ek}.

\section{Acknowledgments}

This work was supported by Basic Science Research Program through
the National Research Foundation of Korea (NRF) funded by the
Ministry of Education (Grant No. 2015R1D1A1A01056572)(P.O.),
the National Research Foundation of Korea(NRF) funded by the Ministry of Education(Grant No. NRF-2017R1D1A1B03032970)(S.K.),
the National Research Foundation of Korea (NRF) funded by the Ministry of Education (Grant No. 2017R1D1A1B03028384)(C.-G.P.).




\begin{thebibliography}{99}

\bibitem{Long:2002wn}
  J.~C.~Long, H.~W.~Chan, A.~B.~Churnside, E.~A.~Gulbis, M.~C.~M.~Varney and J.~C.~Price,
  Nature {\bf 421}, 922 (2003)
  doi:10.1038/nature01432
  [hep-ph/0210004].


\bibitem{Joyce:2014kja}
  A.~Joyce, B.~Jain, J.~Khoury and M.~Trodden,
  Phys.\ Rept.\  {\bf 568}, 1 (2015)
  doi:10.1016/j.physrep.2014.12.002
  [arXiv:1407.0059 [astro-ph.CO]].

\bibitem{Moon:2017rox}
  T.~Moon and P.~Oh,
  JCAP {\bf 1707}, no. 07, 024 (2017)
  doi:10.1088/1475-7516/2017/09/024
  arXiv:1705.00866 [hep-th].


\bibitem{Moon:2009zq}
See  T.~Y.~Moon, J.~Lee and P.~Oh,
  Mod.\ Phys.\ Lett.\ A {\bf 25}, 3129 (2010)
  doi:10.1142/S0217732310034201
  [arXiv:0912.0432 [gr-qc]], and references therein.


\bibitem{Will:2014kxa}
  See, C.~M.~Will,
  Living Rev.\ Rel.\  {\bf 17}, 4 (2014)
  doi:10.12942/lrr-2014-4
  [arXiv:1403.7377 [gr-qc]], and references therein.

\bibitem{Boisseau:2000pr}
  B.~Boisseau, G.~Esposito-Farese, D.~Polarski and A.~A.~Starobinsky,
  Phys.\ Rev.\ Lett.\  {\bf 85}, 2236 (2000)
  doi:10.1103/PhysRevLett.85.2236
  [gr-qc/0001066].

\bibitem{Cooper:1982du} 
  F.~Cooper and G.~Venturi,
  Phys.\ Rev.\ D {\bf 24}, 3338 (1981).
  doi:10.1103/PhysRevD.24.3338
  
\bibitem{Finelli:2007wb}
  F.~Finelli, A.~Tronconi and G.~Venturi,
  Phys.\ Lett.\ B {\bf 659} (2008) 466
  doi:10.1016/j.physletb.2007.11.053
  [arXiv:0710.2741 [astro-ph]].


\bibitem{Umilta:2015cta} 
  C.~Umiltà, M.~Ballardini, F.~Finelli and D.~Paoletti,
  JCAP {\bf 1508}, 017 (2015)
  doi:10.1088/1475-7516/2015/08/017
  [arXiv:1507.00718 [astro-ph.CO]].

\bibitem{Ballardini:2016cvy} 
  M.~Ballardini, F.~Finelli, C.~Umiltà and D.~Paoletti,
  JCAP {\bf 1605}, no. 05, 067 (2016)
  doi:10.1088/1475-7516/2016/05/067
  [arXiv:1601.03387 [astro-ph.CO]].
  
\bibitem{Kamenshchik:2012rs} 
  A.~Y.~Kamenshchik, A.~Tronconi and G.~Venturi,
  Phys.\ Lett.\ B {\bf 713}, 358 (2012)
  doi:10.1016/j.physletb.2012.06.035
  [arXiv:1204.2625 [gr-qc]].
  
\bibitem{Kouwn:2015cdw}
  S.~Kouwn, P.~Oh and C.~G.~Park,
  Phys.\ Rev.\ D {\bf 93}, no. 8, 083012 (2016)
  doi:10.1103/PhysRevD.93.083012
  [arXiv:1512.00541 [astro-ph.CO]].

\bibitem{MCMC}
N. Metropolis, A.W. Rosenbluth, M.N. Rosenbluth, A.H. Teller, and
   E. Teller, 
   J.\ Chem.\ Phys.\ {\bf 21}, 1087 (1953);
W.K. Hastings, 
Biometrika {\bf 57}, 97 (1970).


\bibitem{Hubble:DA}
  C.~Zhang, H.~Zhang, S.~Yuan, T.~J.~Zhang and Y.~C.~Sun,
  Res.\ Astron.\ Astrophys.\  {\bf 14}, no. 10, 1221 (2014)
  [arXiv:1207.4541 [astro-ph.CO]];
  J.~Simon, L.~Verde and R.~Jimenez,
  Phys.\ Rev.\ D {\bf 71}, 123001 (2005)
  [astro-ph/0412269];
  M.~Moresco {\it et al.},
  JCAP {\bf 1208}, 006 (2012)
  [arXiv:1201.3609 [astro-ph.CO]];
  D.~Stern, R.~Jimenez, L.~Verde, M.~Kamionkowski and S.~A.~Stanford,
  JCAP {\bf 1002}, 008 (2010)
  [arXiv:0907.3149 [astro-ph.CO]].
\bibitem{Hubble:BAO}
  C.~H.~Chuang and Y.~Wang,
  Mon.\ Not.\ Roy.\ Astron.\ Soc.\  {\bf 435}, 255 (2013)
  [arXiv:1209.0210 [astro-ph.CO]];
  C.~Blake {\it et al.},
  Mon.\ Not.\ Roy.\ Astron.\ Soc.\  {\bf 425}, 405 (2012)
  [arXiv:1204.3674 [astro-ph.CO]];
  L.~Samushia {\it et al.},
  Mon.\ Not.\ Roy.\ Astron.\ Soc.\  {\bf 429}, 1514 (2013)
  [arXiv:1206.5309 [astro-ph.CO]];
  T.~Delubac {\it et al.} [BOSS Collaboration],
  Astron.\ Astrophys.\  {\bf 574}, A59 (2015)
  [arXiv:1404.1801 [astro-ph.CO]];
  X.~Ding, M.~Biesiada, S.~Cao, Z.~Li and Z.~H.~Zhu,
  Astrophys.\ J.\  {\bf 803}, no. 2, L22 (2015)
  [arXiv:1503.04923 [astro-ph.CO]].

\bibitem{Suzuki:2011hu}
  N.~Suzuki {\it et al.},
  Astrophys.\ J.\  {\bf 746}, 85 (2012)
  [arXiv:1105.3470 [astro-ph.CO]].

\bibitem{SNIa-MRG}
M. Goliath, R. Amanullah, P. Astier, A. Goobar and R. Pain,
  Astron.\ Astrophys.\  {\bf 380}, 6 (2001);
S. Nesseris and L. Perivolaropoulos,
  Phys.\ Rev.\ D {\bf 72}, 123519 (2005).



\bibitem{6dFGS}
 F. Beutler, C. Blake, M. Colless, D.H. Jones, L. Staveley-Smith, L. Campbell, Q. Parker
and W. Saunders {\it et al.},
  Mon.\ Not.\ Roy.\ Astron.\ Soc.\  {\bf 416}, 3017 (2011)
  [arXiv:1106.3366 [astro-ph.CO]].

\bibitem{SDSS}
 N. Padmanabhan, X. Xu, D.J. Eisenstein, R. Scalzo, A.J. Cuesta, K.T. Mehta and E. Kazin,
  Mon.\ Not.\ Roy.\ Astron.\ Soc.\  {\bf 427}, no. 3, 2132 (2012)
  [arXiv:1202.0090 [astro-ph.CO]];
 L. Anderson, E. Aubourg, S. Bailey, D. Bizyaev, M. Blanton, A.S. Bolton, J. Brinkmann
and J.R. Brownstein {\it et al.},
  Mon.\ Not.\ Roy.\ Astron.\ Soc.\  {\bf 427}, no. 4, 3435 (2013)
  [arXiv:1203.6594 [astro-ph.CO]].

\bibitem{WiggleZ}
 C. Blake, S. Brough, M. Colless, C. Contreras, W. Couch, S. Croom, D. Croton and T. Davis
{\it et al.},
  Mon.\ Not.\ Roy.\ Astron.\ Soc.\  {\bf 425}, 405 (2012)
  [arXiv:1204.3674 [astro-ph.CO]].

\bibitem{WMAP9}
 G. Hinshaw {\it et al.}  [WMAP Collaboration],
  Astrophys.\ J.\ Suppl.\  {\bf 208}, 19 (2013)
  [arXiv:1212.5226 [astro-ph.CO]].

\bibitem{Eisenstein-etal-2005}
  D.~J.~Eisenstein {\it et al.}  [SDSS Collaboration],
  Astrophys.\ J.\  {\bf 633}, 560 (2005)
  [astro-ph/0501171].

\bibitem{Eisenstein-Hu-1998}
 D.~J.~Eisenstein and W.~Hu,
  Astrophys.\ J.\  {\bf 496}, 605 (1998)
  [astro-ph/9709112].

\bibitem{Ade:2013zuv}
  P.~A.~R.~Ade {\it et al.} [Planck Collaboration],
  Astron.\ Astrophys.\  {\bf 571}, A16 (2014)
  [arXiv:1303.5076 [astro-ph.CO]].

 \bibitem{Hu-Sugiyama-1996}
 W.~Hu and N.~Sugiyama,
  Astrophys.\ J.\  {\bf 471}, 542 (1996)
  [astro-ph/9510117].

\bibitem{Shafer:2013pxa}
  D.~L.~Shafer and D.~Huterer,
  Phys.\ Rev.\ D {\bf 89}, no. 6, 063510 (2014)
  [arXiv:1312.1688 [astro-ph.CO]].
  
\bibitem{Komatsu:2010fb} 
  E.~Komatsu {\it et al.} [WMAP Collaboration],
  Astrophys.\ J.\ Suppl.\  {\bf 192}, 18 (2011)
  doi:10.1088/0067-0049/192/2/18
  [arXiv:1001.4538 [astro-ph.CO]].



\bibitem{Williams:1995nq}
  J.~G.~Williams, X.~X.~Newhall and J.~O.~Dickey,
  Phys.\ Rev.\ D {\bf 53}, 6730 (1996).
  doi:10.1103/PhysRevD.53.6730


\bibitem{perihelionshift}
  Shapiro I.I.,
  in General Relativity and Gravitation 12,
  Ashby N., et al., Eds. Cambridge University Press (1993).

\bibitem{Shapiro:2004zz}
  S.~S.~Shapiro, J.~L.~Davis, D.~E.~Lebach and J.~S.~Gregory,
  Phys.\ Rev.\ Lett.\  {\bf 92}, 121101 (2004).

\bibitem{Bertotti:2003rm}
  B.~Bertotti, L.~Iess and P.~Tortora,
  Nature {\bf 425}, 374 (2003).

\bibitem{Schimd:2004nq}
  C.~Schimd, J.~P.~Uzan and A.~Riazuelo,
  Phys.\ Rev.\ D {\bf 71}, 083512 (2005)


\bibitem{Muller:2007zzb}
  J.~Muller and L.~Biskupek,
  Class.\ Quant.\ Grav.\  {\bf 24}, 4533 (2007).
  doi:10.1088/0264-9381/24/17/017

\bibitem{Copi:2003xd}
  C.~J.~Copi, A.~N.~Davis and L.~M.~Krauss,
  Phys.\ Rev.\ Lett.\  {\bf 92}, 171301 (2004)
  doi:10.1103/PhysRevLett.92.171301
  [astro-ph/0311334].

\bibitem{Bambi:2005fi}
  C.~Bambi, M.~Giannotti and F.~L.~Villante,
  Phys.\ Rev.\ D {\bf 71}, 123524 (2005)
  doi:10.1103/PhysRevD.71.123524
  [astro-ph/0503502].

\bibitem{Guenther:1998ApJ}
D. B. Guenther, L. M. Krauss, and P. Demarque, Astrophys. J. 498, 871 (1998).

\bibitem{Thorsett:1996fr}
  S.~E.~Thorsett,
  Phys.\ Rev.\ Lett.\  {\bf 77}, 1432 (1996)
  doi:10.1103/PhysRevLett.77.1432
  [astro-ph/9607003].

\bibitem{Hellings:1983PRL}
R. W. Hellings, P. J. Adams, J. D. Anderson, M. S. Keesey, E. L. Lau, E. M. Standish, V. M. Canuto, and I. Goldman,
Phys. Rev. Lett. 51, 1609 (1983).

\bibitem{Kaspi:1994hp}
  V.~M.~Kaspi, J.~H.~Taylor and M.~F.~Ryba,
  Astrophys.\ J.\  {\bf 428}, 713 (1994).
  doi:10.1086/174280

\bibitem{Chan:2007fe}
  K.~C.~Chang and M.-C.~Chu,
  Phys.\ Rev.\ D {\bf 75}, 083521 (2007)
  doi:10.1103/PhysRevD.75.083521
  [astro-ph/0611851].

\bibitem{Wu:2009zb}
  F.~Wu and X.~Chen,
  Phys.\ Rev.\ D {\bf 82}, 083003 (2010)
  doi:10.1103/PhysRevD.82.083003
  [arXiv:0903.0385 [astro-ph.CO]].

\bibitem{Li:2013nwa}
  Y.~C.~Li, F.~Q.~Wu and X.~Chen,
  Phys.\ Rev.\ D {\bf 88}, 084053 (2013)
  doi:10.1103/PhysRevD.88.084053
  [arXiv:1305.0055 [astro-ph.CO]].

\bibitem{Li:2015aug}
  J.~X.~Li, F.~Q.~Wu, Y.~C.~Li, Y.~Gong and X.~L.~Chen,
  Res.\ Astron.\ Astrophys.\  {\bf 15}, no. 12, 2151 (2015)
  doi:10.1088/1674-4527/15/12/003
  [arXiv:1511.05280 [astro-ph.CO]].

\bibitem{LopesFranca:2002ek}
See, for example,   U.~França and R.~Rosenfeld,
  JHEP {\bf 0210}, 015 (2002)
  doi:10.1088/1126-6708/2002/10/015
  [astro-ph/0206194].







\end{thebibliography}
\end{document}